\documentstyle{elsart}

\input{psfig}

\begin{document} 
\begin{frontmatter}

\title{
Notes on the
Kinematic Structure of the Three--Flavor Neutrino Oscillation Framework}

\author{D. V. Ahluwalia}
\footnote{E-mail address: av@p25hp.lanl.gov }
\address{H 846, Physics  Division (P-25) \\
Los Alamos National Laboratory,
Los Alamos, NM 87545, USA}

\maketitle
\begin{abstract}

These notes present a 
 critique of the standard three--flavor neutrino oscillation 
framwork.
The design proposal of the MINOS at Fermilab based
on a two mass eigenstate framework may require serious 
reconsideration 
if there is  strong mixing between all three flavors of neutrinos.
For the 
LSND and KARMEN neutrino 
oscillation experiments,
the amplitude of neutrino oscillation  of the ``one mass scale dominance''
framework
vanishes for certain
values of mixing angles as a result of opposite signs of two equal and opposite
contributions. 
Recent astronomical observations
leave open the possibility that
one of the neutrino mass eigenstates may be non--relativistic
in some instances. 
Neutrino oscillation phenomenology with 
a superposition of  two relativistic, and 
one non--relativistic, mass eigenstates is constructed.
It is concluded that if the transition from the 
non-relativistic to the relativistic regime happens for energies relevant
to the Reactor and the LSND neutrino oscillation experiments then
one must consider an {\em ab intio} analysis of the existing data. 

\end{abstract}
\end{frontmatter}
\section{Introduction}

There are at least two aspects to neutrino oscillation phenomenology.
The first is the kinematic aspect where neutrino oscillations are
modeled  in terms of mass squared differences associated with 
the underlying neutrino mass eigenstates and mixing angles.
The exact form of the standard kinematic analysis depends on whether the
weak neutrino eigenstates are of Dirac type, of Majorana type, or
of the type recently proposed in Ref.
\cite{DVAnp}, and further on whether or not one allows for CP violation in 
the neutrino sector. The second aspect is dynamical in nature.
This includes the well--known MSW modifications to neutrino oscillations
\cite{DR1}, the gravitationally induced neutrino oscillation phases \cite{AB},
and neutrino spin flips in strong gravitational and magnetic fields \cite{PRW}.

In this paper I shall confine my attention to the kinematical aspect
of the phenomenon. 
The neutrino--oscillation parameters determined
by kinematic analysis  may then be used to determine which dynamical aspects
may be important.
It must be noted that even the kinematical aspects of neutrino 
phenomenology \cite{KPbook,BK,MPbook} are only now beginning to be
understood properly  \cite{HL,TG,ABIV}.

As soon as the preprint from  Kamioka reporting the
zenith-angle dependence of the atmospheric neutrino
anomaly \cite{Kamioka} arrived,  it became
clear 
that the phenomenon of neutrino oscillations could no longer be
studied in terms of the two--flavor analysis  {\em and}
that a great amount of confusion had arisen
 when  the intuition and arguments from the 
two--flavor studies were carried over, without due reflection, to a realistic
three--flavor analysis. 
Similar observations were made by Learned, Pakvasa,
and Weiler \cite{tw} 
(and few  others) when they first learned of the 
Kamiokande collaboration's initial
results on  the depletion in the flux of 
atmospheric muon neutrinos relative to the flux of
atmospheric electron neutrinos.
Due to  conceptual and experimental complexities
associated with the neutrino--oscillation physics there are, at present,
a large number of papers  devoted to the 
three--flavor analysis.
Still, some fundamental aspects of the
three--flavor neutrino--oscillation analysis remain to be discussed.
Some of the remarks that I present are seemingly trivial, but in view
of their possible relevance, I take the liberty of presenting them in this 
paper in the hope of
establishing a better conceptual understanding.

In the next section I briefly review the standard three--flavor 
neutrino oscillation framework.  Section 3 is devoted to MINOS at Fermilab
and other long--baseline experiments. Section 4 concentrates on the LSND
neutrino oscillation experiment and KARMEN. Section 5 is confined to a 
brief discussion of the Reactor $\overline{\nu}_e$ experiments. 
Section 6 presents modification to
neutrino--oscillation phenomenology 
with a superposition of relativistic and
non-relativistic mass eigenstates.
\footnote{Section 6 is essentially based on a recent work of the
author with Goldman \cite{DVA21}.}
Section 7 presents the conclusion. In addition to these
seven sections, the paper contains three appendices. 
Appendix A provides further details on  standard neutrino oscillation
phenomenology. 
Appendix B contains a simple theorem on the inverted mass hierarchy.
Appendix C presents a pedagogic discussion of neutrino
oscillations and energy conservation.

\section{Three--Flavor Neutrino Oscillations: A Brief Review}

Because of its simplicity, the two--flavor
neutrino oscillation framwork has dominated  the design and understanding
of neutrino oscillation experiments.
However, as the observational and experimental 
indications for actual neutrino oscillations
mount, newer and conceptually more sophisticated 
neutrino oscillation experiments must be considered.
In these experiments, in my opinion, the insights gained
from the 
two--flavor
neutrino oscillation frameworks may prove to be inadequate.
It is, therefore, important to have a fresh {\em ab initio\/}
look at the three--flavor neutrino oscillations. I begin with a
brief review
of the three--flavor neutrino oscillation scenario and soon
find myself embedded in insights and results that would 
be either counterintuitive from a two--flavor neutrino oscillation
point of view, or entirely absent unless the
three--flavor neutrino oscillations are invoked. Many of these
results arise from a larger number of {\em relative} phases ---
relative phases that can lead, for example, to important cancellations. 

Assuming,  (a) neutrinos to be Dirac particles, that (b)
no CP violation occurs in the neutrino sector, 
that (c) both the flavor eigenstates and the mass eigenstates are
relativistic in the laboratory frame, 
and that (d) all three mass eigenstates are stable,
the kinematically induced
neutrino oscillation probability,
in a three--flavor neutrino oscillation scenario,  reads:
\footnote{When no ambiguity is likely to arise I write
${\cal P}_{\ell\ell^\prime}\left(L, \{\eta_k\}\right)= 
{\cal P}\left(\nu_\ell\rightarrow\nu_{\ell^\prime}\right)$.}
\begin{equation}
{\cal P}_{\ell\ell^\prime}\left(L, \{\eta_k\}\right)= 
 \int_{E_{min}}^{E_{max}} dE\,f_\ell(E)\,
{P}_{\ell\ell^\prime}\left(E, L, \{\eta_k\}\right)\quad,\label{eq:dnew}
\end{equation}
where
\begin{eqnarray}
&&P_{\ell\ell^\prime}\left(E,L,\{\eta_k\}\right)
= 
\delta_{\ell\,\ell'}
-\,\,4\,U_{\ell'\,1}\,U_{\ell\,1}\,U_{\ell'\,2}\,
U_{\ell\,2}\,\sin^2\left(\varphi^0_{2\,1}
\right)\nonumber\\
&&-\,4\,U_{\ell'\,1}\,U_{\ell\,1}\,U_{\ell'\,3}\,
U_{\ell\,3}\,\sin^2\left(
\varphi^0_{3\,1}\right)
\,-\,4\,U_{\ell'\,2}\,U_{\ell\,2}\,U_{\ell'\,3}\,
U_{\ell\,3}\,\sin^2\left(
\varphi^0_{3\,2}\right)\, 
.\label{eq:d}
\end{eqnarray}
The  kinematic phase that appears in the above equation is defined as
\begin{equation}
\varphi^0_{\jmath\imath}
=
2\,\pi \,{L\over\lambda^{\rm osc}_{\jmath\imath }}\quad.
\end{equation}

Equations  (\ref{eq:dnew}) and (\ref{eq:d}) require several additional
comments and observations.
These remarks are enumerated in Appendix A. However, the following
definitions need
to be noted immediately:

\begin{enumerate}

\item
The oscillations length, $\lambda^{\rm osc}_{\jmath\imath }$,
 is defined as
\begin{equation}
\lambda^{{\rm {osc}}}_{\jmath
\imath}\,=\, {{2\,\pi}\over\alpha}\,{E\over{\Delta 
m^2_{\jmath\imath}}}\quad.
\end{equation}
The kinematic phase  may also be written as: 
$\varphi^0_{\jmath\imath}
=1.27 \, \Delta m^2_{\jmath\imath } \times \left({L/ E}\right)$.
These kinematic phases may be modified for dynamical reasons 
\cite{DR1,AB,PRW}.
Here, $\alpha\,=\,\beta/2$; $\beta=2.54$ is the usual factor that
arises from expressing $E$ in $\mbox{MeV}$, $L$ in meters, 
and $\Delta m^2_{j\,k}$ in $\mbox{eV}^2\,$. $E$ refers to neutrino
kinetic 
 energy, $\sqrt{{\vec p}^{\,2}+m^2}$,\footnote{Or perhaps more precisely,
$E$ represents the expectation value of the kinematic Hamiltonian
in the appropriate neutrino--flavor eigenstate.} and
$L$ is the distance between the creation region and the detection region
for the neutrino oscillation event.
The five neutrino oscillation parameters
$\{\eta_k\}$ in  equation (\ref{eq:dnew}) are the two mass squared differences and
the three mixing angles: $\eta_1=\Delta m^2_{2 1}$,
$\eta_2=\Delta m^2_{3 2}$, $\eta_3=\theta$, $\eta_4=\beta$, 
and $\eta_5=\psi$. The third mass squared difference is then given by
$ \Delta m^2_{3 1}=\Delta m^2_{2 1}+\Delta m^2_{3 2}$.

\item
$f_\ell(E)$ is the  neutrino 
flux, of flavor $\ell$, at energy
$E$. The $f_\ell(E)$ is normalized 
to unity:
\begin{equation}
\int_{E_{min}}^{E_{max}} dE \,f_\ell(E) =1\,.
\end{equation}
$E_{min}$ and $E_{max}$ refer, respectively, to the minimum and maximum energy 
relevant to the neutrino beam and the neutrino detector. 

\item
An  element of the $3\times 3$ unitary 
neutrino--mixing--matrix, $U(\theta,\beta,\psi)$, 
is labeled as $U_{\ell j}$ with $\ell$ standing for any of
the three neutrino flavors, $\ell=e,\mu,\tau$; and $j$ representing any of
the three mass eigenstates, $j=1,2,3$.
The explicit expression for the mixing matrix that I use is
(Maiani representation \cite{MR}, with CP phase $\delta$ set equal to zero)
\begin{equation}
U(\theta,\,\beta,\,\psi)\,=\,
\left(
\begin{array}{ccccc}
c_\theta\,c_\beta &{\,\,}& s_\theta\,c_\beta &{\,\,}& s_\beta \\
-\,c_\theta\,s_\beta\,s_\psi\,-\,s_\theta\,c_\psi
&{\,\,}& c_\theta\, c_\psi\,-\,s_\theta\,s_\beta\,s_\psi
&{\,\,}& c_\beta\,s_\psi\\
-\,c_\theta\,s_\beta\,c_\psi\,+\,s_\theta\,s_\psi
&{\,\,}& -\,s_\theta\,s_\beta\,c_\psi\,-\,c_\theta\,s_\psi
&{\,\,}& c_\beta\,c_\psi
\end{array}\right),
\label{eq:umix}
\end{equation}
where $c_\theta\,=\,\cos(\theta)$, $s_\theta\,=\,\sin(\theta)$, etc. 
\end{enumerate}

\section{The $\nu_\mu\rightarrow\nu_\tau$ MINOS at Fermilab, and Other
Long--Baseline Experiments}

For the sake of concreteness let us choose 
a generic long--baseline $\nu_\mu\rightarrow\nu_\tau$ experiment and
consider the mass hierarchy  such that
\begin{equation}
\lambda_{3\,(2,\,1)}^{\rm{osc}} 
\ll L\quad,\quad\quad\lambda^{\rm{osc}}_{2 1} \sim L\quad.\label{scenario}
\end{equation}
Using Eq. (\ref{eq:dnew}) and averaging $\sin^2(\cdots)$ associated
with the $\lambda_{3\,(2,\,1)}^{\rm{osc}}$ terms to
$0.5$ I obtain
\begin{equation}
{\cal P} (\nu_\mu \rightarrow \nu_\tau)\,=\,
A_{\mbox{c}}(\beta,\psi)+\, A_{\mbox{o}}(\theta,\beta,\psi)\,
\int_{E_{min}}^{E_{max}} dE \,f_{\nu_\mu}(E)\,
\sin^2\left( { {C_{2 1}} \over E}\right)\, ,\label{eq:mf}
\end{equation}
where the $\theta$--independent  constant contribution,
$A_{\mbox{c}}(\beta,\psi)$, and the 
amplitude,
$A_{\mbox{o}}(\theta,\beta,\psi)$, of the 
oscillatory term, in 
${\cal P} (\nu_\mu \rightarrow \nu_\tau)$ are:
\begin{eqnarray}
A_{\mbox{c}}(\beta,\psi)\,=\,&&
2\, \cos^4(\beta)\,\cos^2(\psi)\,\sin^2(\psi)
\quad,\\
A_{\mbox{o}}(\theta,\beta,\psi)\,=\,&&
\,4\,\left[\cos(\psi)\,\sin(\theta)\,+\,\sin(\beta)\,\sin(\psi)\,
\cos(\theta)\right] \nonumber\\
&&
\times\,
\left[\sin(\psi)\,\sin(\theta) \,-\,\sin(\beta)\,\cos(\psi)
\,\cos(\theta)\right]\nonumber\\
&&\times\,\left[\sin(\beta)\,\cos(\psi)\,\sin(\theta) \,+\,
\sin(\psi)\,\cos(\theta)\right]\nonumber\\
&&\times\,\left[ \sin(\beta)\,\sin(\psi)\,\sin(\theta)\,-\,
\cos(\psi)\,\cos(\theta)\right]\quad.
\end{eqnarray}

Oftentimes one attempts to understand this experiment, and similar
experiments, within the framework of a two--flavor neutrino oscillation.
Such an analysis can be misleading.
In the physically required three--flavor neutrino oscillations
I see that {\em there is a constant piece in 
${\cal P} (\nu_\mu \rightarrow \nu_\tau)$ and an oscillatory term.
Depending on the mixing angles the constant piece may 
make a significant contribution to the 
${\cal P} (\nu_\mu \rightarrow \nu_\tau)$.}

The counterpart of Eq. (\ref{eq:mf}) for the 
$\nu_e\rightarrow\nu_\tau$ Long--Baseline Experiments reads:
\begin{equation}
{\cal P} (\nu_e \rightarrow \nu_\tau)\,=\,
B_{\mbox{c}}(\beta,\psi)+\, B_{\mbox{o}}(\theta,\beta,\psi)\,
\int_{E_{min}}^{E_{max}} dE \,f_{\nu_e}(E)\,
\sin^2\left( { {C_{2 1}} \over E}\right)\, ,\label{eq:et}
\end{equation}
where 
\begin{eqnarray}
B_{\mbox{c}}(\beta,\psi)\,=\,&&
2\, \cos^2(\beta)\,\sin^2(\beta)\,\cos^2(\psi)
\quad,\\
B_{\mbox{o}}(\theta,\beta,\psi)\,=\,&&
4\,\cos^2(\beta)\,\cos(\theta)\,\sin(\theta) \nonumber\\
&&\times\,\left[\sin(\psi)\,\cos(\theta)
\,+\, \sin(\beta)\,\cos(\psi)\,\sin(\theta) \right]\nonumber\\
&&\times\,\left[
\sin(\psi)\,\sin(\theta)
\,-\, \sin(\beta)\,\cos(\psi)\,\cos(\theta) \right]\quad.
\end{eqnarray}

Similarly, 
the counterpart of Eq. (\ref{eq:mf}) for the 
$\nu_e\rightarrow\nu_\mu$ Long--Baseline Experiments reads:
\begin{equation}
{\cal P} (\nu_e \rightarrow \nu_\mu)\,=\,
C_{\mbox{c}}(\beta,\psi)+\, C_{\mbox{o}}(\theta,\beta,\psi)\,
\int_{E_{min}}^{E_{max}} dE \,f_{\nu_e}(E)\,
\sin^2\left( { {C_{2 1}} \over E}\right)\, ,\label{eq:etb}
\end{equation}
where
\begin{eqnarray}
C_{\mbox{c}}(\beta,\psi)\,=\,&&
2\, \cos^2(\beta)\,\sin^2(\beta)\,\sin^2(\psi)
\quad,\\
C_{\mbox{o}}(\theta,\beta,\psi)\,=\,&&
4\,\cos^2(\beta)\,\cos(\theta)\,\sin(\theta) \nonumber\\
&&\times\,\left[\cos(\psi)\,\sin(\theta)
\,+\, \sin(\beta)\,\sin(\psi)\,\cos(\theta) \right]\nonumber\\
&&\times\,\left[
\cos(\psi)\,\cos(\theta)
\,-\, \sin(\beta)\,\sin(\psi)\,\sin(\theta) \right]\quad.
\end{eqnarray}

It may be worthwhile to observe how, with respect to the 
mixing angle $\psi$, a ``large'' (``small'')
$B_{\mbox{c}}(\beta,\psi)$ 
will complement a ``small'' (``large'')
$C_{\mbox{c}}(\beta,\psi)$.
In particular, {\em by measuring (in the same beam) the  constant pieces in
${\cal P} (\nu_e \rightarrow \nu_\tau)$ and
${\cal P} (\nu_e \rightarrow \nu_\mu)$
I can directly measure the mixing angles $\beta$ and $\psi$:\/}
\begin{eqnarray}
\psi \,=\,\tan^{-1}\left[\sqrt{ {{C_{\mbox{c}}(\beta,\psi)}\over
{B_{\mbox{c}}(\beta,\psi)}} }\right]\quad,\label{psi} \\
\beta\,=\,{1\over 2}\,
\sin^{-1}\left({{\sqrt{2\,B_{\mbox{c}}(\beta,\psi)}}
\over{\cos(\psi)}}\right)\,=\,
{1\over 2}\,
\sin^{-1}\left({{\sqrt{2\,C_{\mbox{c}}(\beta,\psi)}}
\over{\sin(\psi)}}\right)\quad.\label{beta}
\end{eqnarray}
In Eq. (\ref{beta}) the angle $\psi$ is to be substituted from
Eq. (\ref{psi}).

Hence, 
the angle $\theta$, the remaining of the three mixing angles,
can be obtained by measuring the amplitude of the oscillatory
term in ${\cal P} (\nu_\mu \rightarrow \nu_\mu)$,
or equivalently  any one of the 
$A_{\mbox{o}}(\theta,\beta,\psi)$, $B_{\mbox{o}}(\theta,\beta,\psi)$,
$C_{\mbox{o}}(\theta,\beta,\psi)$.
Within the scenario represented by Eq. (\ref{scenario}),
the three probabilities (see comments below that require
an additional measurement)
 for
$\nu_\mu\rightarrow\nu_\tau$,
$\nu_e\rightarrow\nu_\tau$, and $\nu_e\rightarrow\nu_\mu$, 
measure four parameters  ---
the three mixing angles $(\theta, \beta, \psi)$ and one
of the two independent mass square differences $\Delta m^2_{2 1}$.
The long--baseline 
experiments, by their very nature (within the scenario under consideration),
determine only $\Delta m^2_{2 1}$. The $\Delta m^2_{3 2}$ must then
be determined either in a reactor experiment or an experiment like
the LSND neutrino oscillation experiment.

In principle, the three
${\cal P}(\nu_\ell\rightarrow\nu_{\ell^\prime})$
considered above can be measured at a very similar $f_{\nu_\ell}(E)$.
However, an additional measurement of one of the three
${\cal P}(\nu_\ell\rightarrow\nu_{\ell^\prime})$
must be done at a sufficiently different $f_{\nu_\ell}(E)$.
This becomes obvious if I note 
 that there  are nine oscillation probabilities,
\begin{equation}
\begin{array}{ccccc}
{\cal P}(\nu_e\rightarrow\nu_e), &\quad&
{\cal P}(\nu_e\rightarrow\nu_\mu), &\quad&
{\cal P}(\nu_e\rightarrow\nu_\tau), \\
{\cal P}(\nu_\mu\rightarrow\nu_e), &\quad&
{\cal P}(\nu_\mu\rightarrow\nu_\mu), &\quad&
{\cal P}(\nu_\mu\rightarrow\nu_\tau), \\
{\cal P}(\nu_\tau\rightarrow\nu_e), &\quad&
{\cal P}(\nu_\tau\rightarrow\nu_\mu), &\quad&
{\cal P}(\nu_\tau\rightarrow\nu_\tau), \label{x1}
\end{array}
\end{equation}
that can be measured at a given energy $E$.
Because of the three unitarity--imposed conditions,
\begin{equation}
{\cal P}(\nu_\ell\rightarrow\nu_e) \,+\,
{\cal P}(\nu_\ell\rightarrow\nu_\mu) \,+\,
{\cal P}(\nu_\ell\rightarrow\nu_\tau) \,=\,1 , \,\,\ell=e,\mu,\tau\quad,
\label{x2}
\end{equation}
and the three conditions 
\begin{eqnarray}
{\cal P}(\nu_e\rightarrow\nu_\mu) =&&
{\cal P}(\nu_\mu\rightarrow\nu_e)\quad,\nonumber\\
{\cal P}(\nu_e\rightarrow\nu_\tau) =&&
{\cal P}(\nu_\tau\rightarrow\nu_e)\quad,\nonumber\\
{\cal P}(\nu_\mu\rightarrow\nu_\tau) =&&
{\cal P}(\nu_\tau\rightarrow\nu_\mu)\quad,\label{x3}
\end{eqnarray}
arising from the tentative assumption that
there is no CP violation in the neutrino sector,
only three of the ${\cal P}(\nu_\ell\rightarrow\nu_{\ell^\prime})[f(E)]$ are 
independent. Therefore, to determine the four parameters
$(\theta, \beta,\psi,\,\, \Delta m^2_{2 1})$ one needs
one more measurement
${\cal P}(\nu_\ell\rightarrow\nu_{\ell^\prime})[f(E^\prime)\ne f(E)]$.
\footnote{ In Eqs. (\ref{x1}) to (\ref{x3}) all
${\cal P}(\nu_\ell\rightarrow\nu_\ell^\prime)$ 
refer to same energy spectrum $f(E)$.}

\section{Experiments Dedicated
to $\overline{\nu}_\mu \rightarrow \overline{\nu}_e$
in the Appearance Mode: LSND and KARMEN}

The published results  of the 
LSND
Neutrino Oscillation Experiment (NOE)
and Karlsruhe--Rutherford Medium Energy Neutrino Experiment
(KARMEN) both present
results on
$\overline{\nu}_\mu \rightarrow \overline{\nu}_e$ in the 
appearance (of $\overline{\nu}_e$) mode \cite{LSND,KARMEN}.  
For LSND NOE  $L \simeq 30\,\, \mbox{m}$
and for KARMEN $L \simeq 17\,\, \mbox{m}$. KARMEN, which is less sensitive
than LSND NOE, sees no neutrino--oscillation signal.
The latest LSND NOE results,
if interpreted within neutrino oscillation framework,
yield  
\[{\cal P}(\overline{\nu}_\mu\rightarrow\overline{\nu}_e)\vert_{LSND}\,=\,
\left[0.31^{+0.11}_{-0.10}\pm 0.05\right]\times 10^{-2}.\]

For the $\mu^+$ decay at  rest, the normalized--to--unity
 Michel energy spectrum
of the $\overline{\nu}_\mu$
is given by
\begin{equation}
f_{\overline{\nu}_\mu}(y)\,=\,
-\,{{2 y_2^3}\over{y^4}}\left(3\,-\,{{2\,y_2}\over y}\right)\quad,
\label{eq:michel}
\end{equation}
where $y\equiv 1/E$, and 
\begin{equation}
y_2\,=\,{1\over{E_{max}}}\,\,;\quad E_{max}=52.8\,\,{\mbox{MeV}}\quad.
\end{equation}

Substituting Eq. (\ref{eq:michel}) into Eq. (\ref{eq:dnew}) I obtain:

\vbox{
\begin{eqnarray}
{\cal P}(\overline{\nu}_\mu\rightarrow\overline{\nu}_e)
\,=\, 
\,-\,&&  4\,U_{e\,1}\,U_{\mu\,1}\,U_{e\,2}\,
U_{\mu\,2}\, \times\, {\cal O}(L,\,\Delta m^2_{2\,1})\nonumber\\
\,-\,&& 4\,U_{e\,1}\,U_{\mu\,1}\,U_{e\,3}\,
U_{\mu\,3}\, \times\,
{\cal O}(L,\,\Delta m^2_{3\,1}) \nonumber\\
\,-\,&& 4\,U_{e\,2}\,U_{\mu\,2}\,U_{e\,3}\,
U_{\mu\,3}\, \times\,
{\cal O}(L,\,\Delta m^2_{3\,2})\, 
\quad.\label{eq:lk}
\end{eqnarray}
}	
In the above expression I have introduced a new function
${\cal O}(L,\,\Delta m^2_{\jmath\imath})$. It is defined as:

\begin{eqnarray}
{\cal O}&&(L,\,\Delta m^2_{\jmath\imath}) =
 2\,y_2^3{\Bigg[}
\left({{C_{\jmath\imath} }
\over {3\,y_2^2}}
+
{{C_{\jmath\imath}^3}
\over
{3}} \right)
\sin\left(2\,y_2\,C_{\jmath\imath}\right)
+
\left({{5\,C_{\jmath\imath}^5}
\over
{6\,y_2}}
-{1\over {4\,y_2^3}}\right)
\cos\left(2\,y_2\,C_{\jmath\imath}\right)\nonumber\\
&&\quad\quad+\,
2\,C_{\jmath\imath}^3\,\mbox{Si}\left(2\,y_2\,C_{\jmath\imath}\right)
-
{{2\,y_2\,C_{\jmath\imath}^4}
\over
{3}}\,
\mbox{Ci}\left(2\,y_2\,C_{\jmath\imath}\right)
+{1\over{4\,y_2^3}}
-\pi C_{\jmath\imath}^3 {\Biggr]}\quad,
\end{eqnarray}
where  $C_{\jmath\imath}=\alpha L \Delta m^2_{\jmath\imath}\,$ and 
\begin{equation}
{\mbox{Si}}(x)\,\equiv\,\int_0^x dt { {\sin(t)}\over t }\,,\quad
{\mbox{Ci}}(x)\,\equiv\,\int_0^x dt { {\cos(t)}\over t }\quad.
\end{equation} 
Given the mixing angles and detector efficiencies
(that can depend not only on $E$ but also on the location of an event
within the detector),
the function ${\cal O}(L,\,\Delta m^2)$ is a measure of
the neutrino oscillation probability.
 I shall call ${\cal O}(L, \,\Delta m^2)$ the  {\em probability function}. 
The probability function, corresponding
to LSND's $L\simeq 30\,\,\mbox{m}$, 
is graphed in Fig. 1.

As it should, ${\cal O}(L,\,\Delta m^2)$ approaches
$0.5$ as $\Delta m^2 \rightarrow\infty$.

In an experiment
using the neutrinos from the decay at rest of $\mu^+$
(and in the absence of any directed magnetic fields),
the event density within the detector 
is measured by the  {\em raw 
event density} function, ${\cal E}(L,\,\Delta m^2)$,
\begin{eqnarray}
{\cal E}(L,\,\Delta m^2) \equiv&& {\epsilon(L)\over {L^2}}\,\times\,
{\cal P}\left(\overline{\nu}_\mu\rightarrow\overline{\nu}_e\right)
\quad,\nonumber\\
\equiv&&\epsilon(L)\,{\cal E}^\prime(L,\,\Delta m^2)\quad, 
\label{eq:ed}
\end{eqnarray}
where $\Delta m^2=\{\Delta m^2_{2\,1},\,\Delta m^2_{3\,2}\}$ and
$\epsilon(L)$ is the  energy--integrated average efficiency at $L$ inside 
the
detector, and the definition of 
${\cal E}^\prime(L,\,\Delta m^2)$, the {\em event density} function, is
obvious from Eq. (\ref{eq:ed}).
In actual experimental situations, such as LSND's NOE and KARMEN,
the gain in event rate with decreasing $L$ is inevitably
shadowed by a increased background \cite{BL}. By measuring the 
$L$--dependence of the 
event 
density function within a detector one may rule
out one set of solutions over the other.

To explore the physical content of Eq. (\ref{eq:lk}) for
${\cal P}(\overline{\nu}_\mu\rightarrow\overline{\nu}_e)$
let us consider an often--invoked scenario in which
mass eigenstates
$\vert \nu_1\rangle $ and  $\vert \nu_2\rangle $ are almost
degenerate \cite{Fogli}. 
Setting 
$\Delta m^2_{3\,2} \simeq \Delta m^2_{3\,1} \equiv
\Delta m^2_{3\,(2,1)}$, and using Eq. (\ref{eq:umix}),
Eq. (\ref{eq:lk}) becomes:
\begin{equation}
{\cal P}(\overline{\nu}_\mu\rightarrow\overline{\nu}_e)
\,=\,{\cal A}_{2\,1}\,
{\cal O}(L,\,\Delta m^2_{2\,1})
\,+\,{\cal A}_{3(2,1)}\,{\cal O}(L,\,\Delta m^2_{3\,(2,1)})
\quad,\label{eq:omd}
\end{equation}
with the two oscillation amplitudes defined as
\begin{eqnarray}
{\cal A}_{2\,1}&&=
4\,\cos^2(\beta)\,\cos(\theta)\,\sin(\theta)
\,{\Bigl[}\cos(\psi)\,\sin(\theta) \,+\,
\sin(\beta)\,\sin(\psi)\,\cos(\theta){\Bigr]}\nonumber\\
&&\times
{\Bigl[}
\cos(\psi)\,\cos(\theta) \,-\,
\sin(\beta)\,\sin(\psi)\,\sin(\theta)
{\Bigr]}
\quad,\nonumber\\
{\cal A}_{3(2,1)}&&=4\,\cos^2(\beta)\,\sin^2(\beta)\,\sin^2(\psi)\quad.
\label{eq:zzz}
\end{eqnarray}

In the scenario indicated above one usually assumes
$m_3\gg m_2\simeq m_1$.
For $\Delta m^2_{2\,1} \ll \Delta m^2_{3\,(2,1)}$, and based
on the $\Delta m^2$--dependence of $\sin^2(y\,C_{\jmath\imath})$,
one neglects the term associated with $\Delta m^2_{2\,1}$ and assumes
that the dominant contribution to 
${\cal P}(\overline{\nu}_\mu\rightarrow\overline{\nu}_e)$
comes from the $\Delta m^2_{3\,(2,1)}$ term.
This is the basic scenario that the one--mass scale dominance
framework considers.

I now make two observations in this context.
\begin{description}

\item[Observation I]
For 
$\beta=
n\,\pi/2$
and/or  $\psi=n\,\pi$, $n=0,1,2, \ldots$,
the oscillation amplitude \footnote{For ``and'' in ``and/or'' above
the three--flavor analysis reduces to a two--flavor framework.}
${\cal A}_{3(2,1)}$
associated with the 
${\cal O}(L,\,\Delta m^2_{3\,(2,1)})$ term,
the dominant term of the so called ``one mass--scale dominance''
 identically vanishes.

\item[Observation II]
Refer to Fig. 1, and consider a set of mass--squared differences such that
$\Delta m^2 _{2,\, 1}$ lies roughly between $1$ and $2$ $\mbox{eV}^2$ and 
$\Delta m^2 _{3(2,\, 1)}$ is greater than $2$ $\mbox{eV}^2$. Then
\begin{equation}
{\cal O}\left(L,\,\Delta m^2_{2,\,1}\right)
 \,\ge\, {\cal O}\left(L,\,\Delta m^2_{3(2,\, 1)} \right)
\quad.\label{eq:v}
\end{equation}
For the above configuration of mass--squared differences,
Eq. (\ref{eq:v})
 violates one of the 
 assumptions of the ``one mass--scale dominance.''
That is,   the 
$\ge$ in the above expression 
is in contradiction to the expected $\ll$ of the 
the one mass--scale dominance analysis. \footnote{Note should be taken that
specific values of $\Delta m^2$ considered here refer to $L=30$ meters
for LSND NOE. 
In addition, as  the detector has a width of about $8$ meters centered 
at $L=30$ meters, depending on  
the oscillation length under consideration, appropriate $L$--integration 
of Eq. (\ref{eq:v}) may need to be considered.
Similar considerations apply to KARMEN.}

\end{description}

Observations I and II suggest that 
the ${\cal O}(L,\,\Delta m^2_{2\,1})$ term may make a
dominant contribution to
${\cal P}(\overline{\nu}_\mu\rightarrow\overline{\nu}_e)$ for certain
values of $\beta$, $\psi$, and $\Delta m^2_{2\,1}$.

\begin{center}
\centerline{\psfig{figure=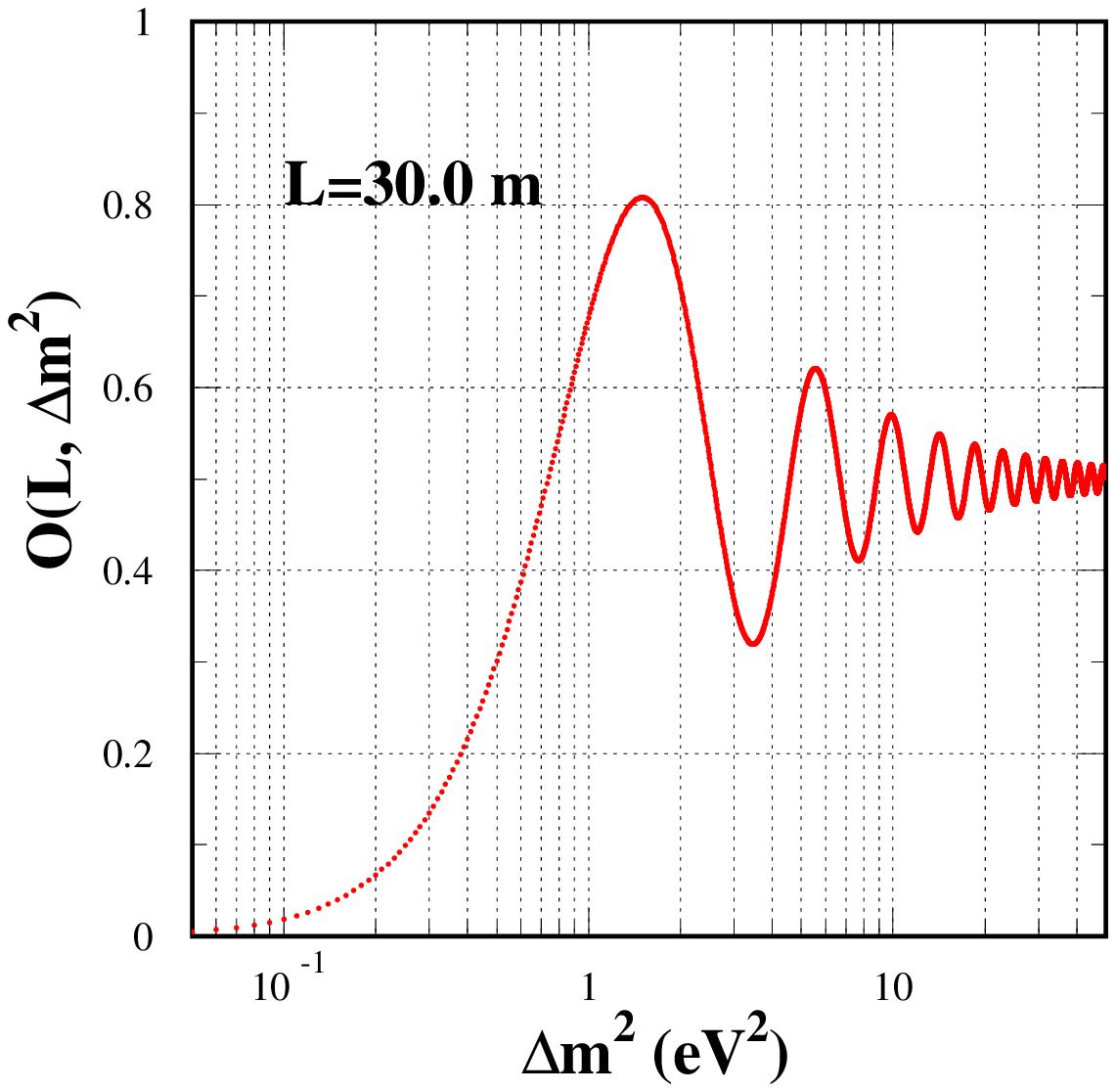,height=12.0cm}}

\vspace{-0.5in}
\begin{minipage}{10.0cm} {\small{{\bf Figure 1.} {\em
The function ${\cal O}(L,\,\Delta m^2)$, corresponding
to LSND's $L\simeq 30\,\,\mbox{m}$.}}}
\end{minipage}
\end{center}

For the indicated--example configuration of mass--squared differences,
the canonical wisdom of the one--mass scale dominance fails, in
part, because
${\cal P}(\overline{\nu}_\mu\rightarrow\overline{\nu}_e)$
evaluated
at 
\begin{equation}
E_{\rm average} \equiv 
\int_\infty^{y_2} dy \,{{f_{\overline{\nu}_\mu}(y) }\over y}\quad,
\end{equation}
does not equal Michel--spectrum averaged
${\cal P}(\overline{\nu}_\mu\rightarrow\overline{\nu}_e)$.
This is particularly true for low $\Delta m^2$, that is, 
{\em before\/} the $\Delta m^2$--region where ${\cal O}(L,\,\Delta m^2)$
reaches its $(\Delta m^2 \rightarrow \infty)$--value of one half
[which in turn coincides with  the average value 
 of $\sin^2(\cdots)$)].

The above discussion should not be interpreted  to mean that
the existing neutrino--oscillation data  necessarily
imply the specific values of parameters for which the 
one--mass scale dominance requires revision. It remains possible
that actual neutrino--oscillation parameters satisfy the canonical 
one--mass scale analysis. In fact, the LSND NOE relevant $\Delta m^2$
of Ref. \cite{Fogli} lies within the applicability of one--mass
scale dominance analysis as will be shown elsewhere.
However, it is important to define
the unexpected boundaries where the  
one--mass scale dominance analysis can, and does, break down.

\begin{center}
\centerline{\psfig{figure=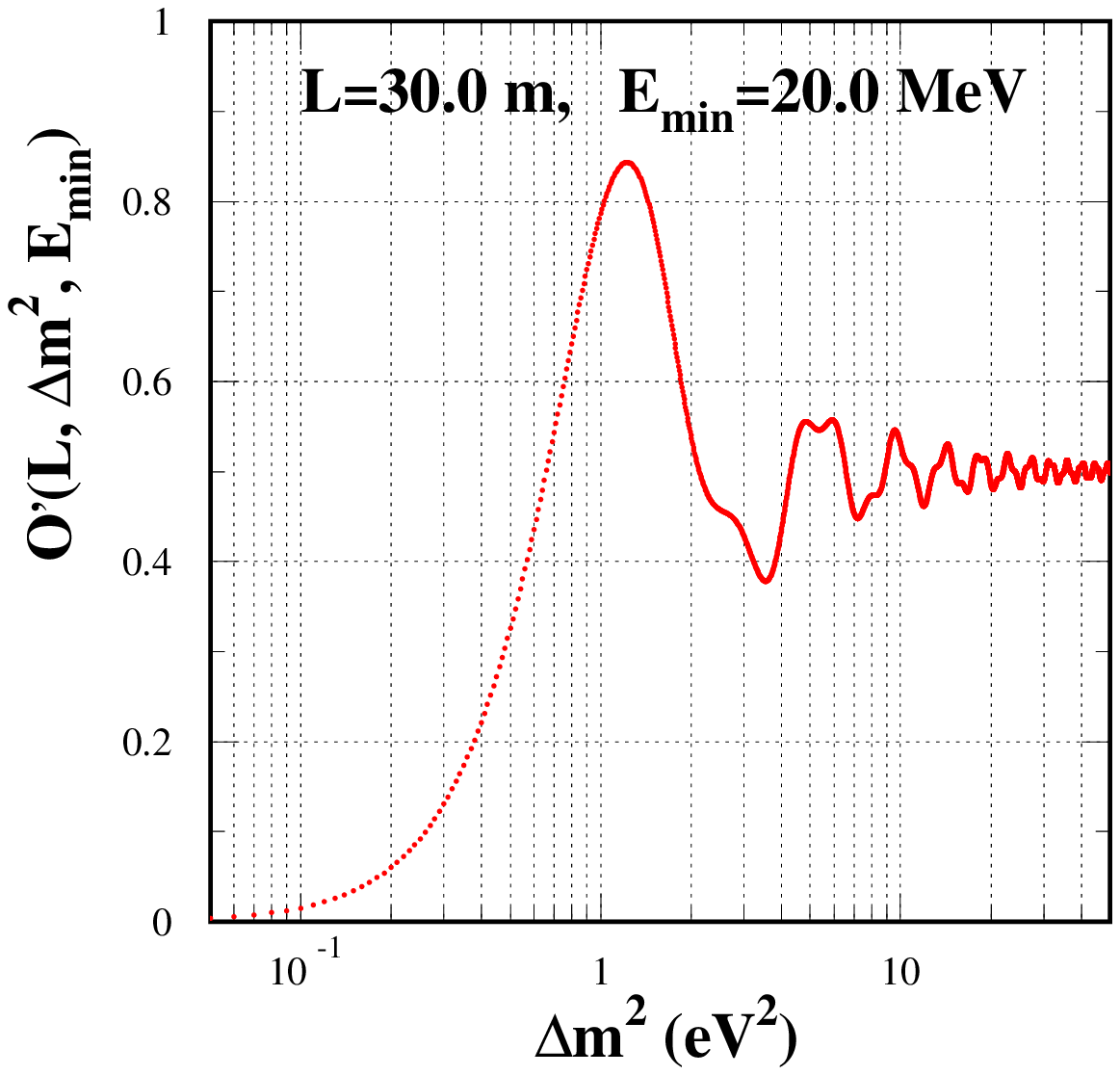,height=12.0cm}}
\vspace{-0.5in}
\begin{minipage}{10.0cm} {\small{{\bf Figure 2.} {\em
The  functions
${\cal O}^\prime(L,\,\Delta m^2_{\jmath\imath},\,E_{min})$ for LSND NOE.}}}
\end{minipage}
\end{center}

For various reasons, such as detector efficiency, 
neutrino background, etc. \cite{LSND,KARMEN}, one may wish
to introduce an energy cutoff in $E_\nu$. 
For LSND's decay--at--rest data this energy cutoff corresponds to 
$E_{min}\simeq 20\,\,{\mbox{MeV}}$.
Under such circumstances, assuming that there is no cut 
on $E_{max}$ except that dictated by the Michel spectrum itself,
the 
${\cal O}(L,\,\Delta m^2_{\jmath\imath})$ in the above discussion should be 
replaced  by
\begin{eqnarray}
{\cal O}^\prime 
(L,\,\Delta m^2_{\jmath\imath},\,E_{min}) =&&
\left(
{1\over {4 \,y_2}} - {{4 \,y_2^2 -12\, y_1\, y_2 + 9\, y_1^2}
\over{4\, y_1^2\, y_2}}
\right)^{- 1} 
{\Biggl[} {3\over{2\, y_1}} - {{y_2}\over {2\, y_1^2}} - {1\over {y_2}}
\nonumber\\
&&+\, C_{\jmath\imath}\,\sin\left(2\,y_2\,C_{\jmath\imath}\right)
- {{C_{\jmath\imath}\, y_2}\over {y_1}}
\sin\left(2\,y_1\,C_{\jmath\imath}\right) \nonumber\\
&&
+\, {1\over {y_2}} \cos\left(2\,y_2\,C_{\jmath\imath}\right)
+ {{y_2}\over {2\, y_1^2}} 
\cos\left(2\,y_1\,C_{\jmath\imath}\right)
- {3\over{2\, y_1}}
\cos\left(2\,y_1\,C_{\jmath\imath}\right)\nonumber\\
&&
+\, 3\, C_{\jmath\imath}\,\mbox{Si}\left(2\,y_2\,C_{\jmath\imath}\right)
- 3\, C_{\jmath\imath}\,\mbox{Si}\left(2\,y_1\,C_{\jmath\imath}\right)
\nonumber \\
&& +\, 2\, y_2\, 
C^2_{\jmath\imath}\,\mbox{Ci}\left(2\,y_1\,C_{\jmath\imath}\right)
- 2\, y_2 \,C^2_{\jmath\imath}\,\mbox{Ci}\left(2\,y_2\,C_{\jmath\imath}\right)
{\Biggr]}
\quad,\label{ojiprime}
\end{eqnarray}
where $y_1 = 1/E_{min}$. 
The new function
${\cal O}^\prime(L,\,\Delta m^2_{\jmath\imath},\,E_{min})$ is 
plotted in Fig. 2 for LSND. 
The differences and similarities between the  functions
${\cal O}(L,\,\Delta m^2_{\jmath\imath},\,E_{min})$ and
${\cal O}^\prime(L,\,\Delta m^2_{\jmath\imath})$ are 
apparent in Fig. 2.

\section{Experiments Dedicated to 
$\overline{\nu}_e \rightarrow
\overline{\nu}_e $ in the Disappearance Mode: Reactor
$\overline{\nu}_e$ Experiments }

These experiments are performed with reactor $\overline{\nu}_e$,
which have a typical energy of about $5\,\,\mbox{MeV}$.
For this section I shall assume $E = 5\,\,\mbox{MeV}$.
In extracting the five neutrino--oscillations parameters 
from the
existing data we must use the exact known spectra and detector
efficiencies.

At the present time there exist at least four results
 on
${\cal P}(\overline{\nu}_e \rightarrow \overline{\nu}_e)$
from reactor experiments  \cite{R1,R2,R3,R4}.
These experiments work in the 
$\overline{\nu}_e$ disappearance mode. The core and the  reactor sizes
place  a lower limit on the distance between the core and the detector.
For the latest experiment from Savannah River we have 
$L$ of about $18\,\,\mbox{m}$, for one position of the
detector, and  roughly $24\,\,\mbox{m}$ for another. 
The overall best limit on 
${\cal P}(\overline{\nu}_e \rightarrow \overline{\nu}_e)$ comes from
the Bugey experiment \cite{R3}.
Results from all reactor experiments are consistent with the 
null--oscillation hypothesis. The oscillation hypothesis, within 
the framework of a 
two--flavor neutrino oscillation, for the 
Savannah River experiment \cite{R4} yields  
$\Delta m^2=3.84\,\,\mbox{eV}^2$. The existing cosmological arguments
suggest that $\sum_i m_i \le 30 \,\,\mbox{eV}$. Thus, 
the range $0 \,\,\mbox{eV}^2 
\le \Delta m^2_{\jmath\imath}\le 10^3\,\,\mbox{eV}^2$ 
holds  particular interest.

There are two independent $\Delta m^2_{\jmath\imath}$. Let these be
$\Delta m^2_{2\,1}$ and $\Delta m^2_{3\,2}$. Then, the third
$\Delta m^2_{\jmath\imath}$ is simply given by: 
$\Delta m^2_{3\,1}=\Delta m^2_{2\,1} + \Delta m^2_{3\,2}$.
Associated with these $\Delta m^2_{\jmath\imath}$ are three
oscillation lengths (of which only two are independent): 
$\lambda^{\rm osc}_{2\,1}$,
$\lambda^{\rm osc}_{3\,2}$, and 
\begin{equation}
\lambda^{\rm osc}_{3\,1} = 
{{\lambda^{\rm osc}_{2\,1}\,\lambda^{\rm osc}_{3\,2}}
\over
{\lambda^{\rm osc}_{2\,1}+\lambda^{\rm osc}_{3\,2}}}\quad.
\end{equation}
The initial three independent length scales associated with
the masses $m_\imath$ of the mass eigenstates $\vert\nu_\imath\rangle$
reduce to two independent length scales. The physical origin of this
fact lies in the assumption that all three mass eigenstates are
relativistic.

To gain some understanding of the 
neutrino--oscillation parameter space\\
$\left(\Delta m_{21}^2,
\Delta m_{32}^2,\theta,\beta,\psi\right)$ 
suppose that 
$\Delta m^2_{2\,1} \simeq \Delta m^2_{3\,2} \simeq 10^3\,\,\mbox{eV}^2$.
Then, 
$\lambda^{\rm osc}_{2\,1} \simeq \lambda^{\rm osc}_{3\,2} \simeq
3\,\,\mbox{cm}$ and a detector placed at $L\gg 3\,\,\mbox{cm}$ sees only 
an overall $\overline{\nu}_e$ deficit:
\begin{equation}
{\cal P}(\overline\nu_e\rightarrow\overline\nu_e) =
1-2 \cos^2(\beta)\left(\sin^2(\beta)+\cos^2(\beta)\cos^2(\theta)\sin^2(\theta) 
 \right). \label{eq:sun}
\end{equation}
This expression follows 
by setting all $\sin^2(\cdots)$ in Eq. (\ref{eq:dnew}) equal to
$0.5$,
as is appropriate when the corresponding $\lambda^{\rm{osc}}\gg $ the
relevant $L$,
 setting
$\ell=\ell^\prime=e$, and
exploiting certain trigonometric identities.

For this scenario, the reactor experiment is unable to determine any of
the $\Delta m^2_{\jmath\imath}$. It is only sensitive to the two
of the three mixing angles. The $\overline{\nu}_e$ deficit, if
$\Delta m^2_{2\,1} \simeq \Delta m^2_{3\,2} \simeq 10^3\,\,\mbox{eV}^2$,
in reactor experiments is  independent of the mixing angle
$\psi$. The Savannah River experiment yields (in the ``integrated rate
test'')
${\cal P}(\overline\nu_e\rightarrow\overline\nu_e) =
\left[98.7 \pm 0.6 \,\,\mbox(stat.) \pm 3.7\,\,\mbox(syst.)\right]
\times 10^{-2}$ for ``position 1'' of the detector and
${\cal P}(\overline\nu_e\rightarrow\overline\nu_e) =
\left[105.5 \pm 1.0 \,\,\mbox(stat.) \pm 3.7\,\,\mbox(syst.)\right]
\times 10^{-2}$ for ``position 2'' of the detector. These rates
are in good agreement with the no--oscillation hypothesis.

Since,
in the above defined scenario,
 the condition $L\gg 3\,\,\mbox{cm}$ for the Sun  is satisfied by
thirteen orders of magnitude, and if
CP is not violated in the neutrino sector, and a solar neutrino deficit arises
solely from a kinematically induced neutrino oscillation, then
Eq. (\ref{eq:sun}) is valid for Sun as well. That is, in the
above defined scenario,
${\cal P}(\overline\nu_e\rightarrow\overline\nu_e)\vert_{\rm Reactor} =
{\cal P}(\nu_e\rightarrow\nu_e)\vert_{\rm Solar} $. But since \cite{Solar}
${\cal P}(\nu_e\rightarrow\nu_e)\vert_{\rm Solar} \sim 0.5 $, {\em the scenario}
 $\Delta m^2_{2\,1} \simeq \Delta m^2_{3\,2} \simeq 10^3\,\,{\mbox{eV}}^2$
is ruled out.
{\em So, 
at least one of the} $\Delta m_{\jmath\imath}^2 \ll 10^3\,\,{\mbox{eV}}^2$.

This is a generally known conclusion but, to the best of our knowledge, 
here it has been reached via
 an entirely new chain of arguments.

\section{Neutrino Oscillations with a Superposition of Relativistic and
Non-Relativistic Mass Eigenstates} 

The existing indications of neutrino oscillations \cite{Solar,LSND,Kamioka}
 arise from the 
data that contains neutrino energies in sub-MeV to GeV range.
The neutrino oscillation phenomenology within the standard three--flavor
framework contains the fundamental assumption that neutrino mass
eigenstates that superimpose to yield the neutrino flavor eigenstates
are relativistic.
This assumption is supported by the cosmological argument \cite{GB}  that the
sum of the neutrino masses have an upper bound of about $30\,\,\mbox{eV}$
\begin{equation}
\sum_\ell m(\nu_\ell)\,\le\, 30 \,\,\mbox{eV}\quad.\label{eq:con}
\end{equation}
However, very recent astronomical observations \cite{news1,pc1,news2}
raise potentially 
serious questions on the validity of the standard
cosmological model. First \cite{news1,pc1}, the UC Berkeley's
Extreme Ultraviolet Explorer  satellite's
observations of the Coma cluster of galaxies indicates that 
this cluster of galaxies may contain a submegakelvin cloud of $\sim 10^{13}
M_\odot$ baryonic gas. Second \cite{news2}, the discovery by the
German x-ray satellite Rosat in which
a sample of 24 Seyfert galaxies contained 12 that were accompanied by a 
pairs of x--ray sources, almost certain to be high redshift quasars,
aligned on either side of the galaxy.
These observations and the associated interpretations, if correct, may 
place
severe questions to cosmological models that depend on the ratio
of photon to baryonic density in the
universe and its size.

Tentatively, therefore, I relax the
cosmological constraint ({\ref{eq:con}) completely and explore
the resulting consequences for three-flavor neutrino osccillation 
framework.
The latest kinematic limits on neutrino masses 
are much less severe
\cite{taumass,PDB}:
\begin{eqnarray}
&&m(\nu_\tau) < 23 \,\,\mbox{MeV}\quad,\label{eq:tmass}\\
&&m(\nu_\mu) < 0.17 \,\,\mbox{MeV}\quad,\label{eq:mmass}\\
&&m(\nu_e) < 10-15\,\,\mbox{eV}\quad. \label{eq:emass}
\end{eqnarray}
The existing and the proposed neutrino oscillation experiments
involve neutrino energies from a fraction of a MeV, if not less,
to several hundred  $\mbox{GeV}$
for the upper--energy end of the neutrino beams.
It remains possible that for some of the experiments
(or a certain sector of an experiment) the mass eigenstates 
$\vert\nu_3\rangle$ and $\vert\nu_2\rangle$
are non relativistic.
Towards the end of understanding such a possible situation
I now consider the 
interplay of non-relativistic and relativistic
mass eigenstates in three--flavor neutrino oscillation framework.

Consider a physical situation where 
we have two relativistic, $\vert\nu_1\rangle$ and
$\vert\nu_2\rangle$,  and one 
non--relativistic, $\vert\nu_3\rangle$,  neutrino mass eigenstates. 

At $t=0$, $x=0$, assume that a source  creates
a $\nu_\ell$
\begin{equation}
\vert\nu_\ell\rangle\,=\,U_{\ell 1}\,\vert\nu_1\rangle
\,+\,U_{\ell 2}\,\vert\nu_2\rangle
\,+\,U_{\ell 3}\,\vert\nu_3\rangle\quad.
\end{equation}
The spatial envelope, which is assumed to be ``sufficiently'' narrow, 
associated with $\{\vert\nu_1\rangle,\,\vert\nu_2\rangle\}$ evolves
towards the detector as $x\simeq t$, while the spatial envelope
of $\vert\nu_3\rangle$ evolves towards the detector as 
\begin{equation}
x\simeq (p/{m_3})\,t = \left({{\sqrt{2 m_3(E-m_3)}}\over
{m_3}}\right) t\quad.
\end{equation}
Therefore, the 
$\{\vert\nu_1\rangle,\,\vert\nu_2\rangle\}$ arrives at the 
detector at time $t_I\simeq L$, while the
$\vert\nu_3\rangle$ arrives at the detector at 
a time $t_{II} = m_3 L/\sqrt{2 m_3(E-m_3)}$. For the above considered energies and
for a source--detector distance of a few tens (or greater) of meters the 
detected $\vert\nu_{\ell^\prime}\rangle$ has no overlap with
$\vert\nu_3\rangle$ at the registered event at $t_I$, and similarly
the detected $\vert\nu_{\ell^\prime}\rangle$ has no overlap with
$\{\vert\nu_1\rangle,\,\vert\nu_2\rangle\}$ at the registered event at $t_{II}$.
With these observations at hand one can easily evaluate the modification
to the neutrino oscillation probability (\ref{eq:d}). The modified
expression reads:
\begin{eqnarray}
P_{\ell\ell^\prime}\left(E,L,\{\xi_k\}\right)
= 
\left(U_{\ell^\prime 3}\,U_{\ell 3}\right)^2 + && {\Big[} \left(
U_{\ell^\prime 1}\,U_{\ell 1}
\,+\,
U_{\ell^\prime 2}\,U_{\ell 2}\right)^2 \nonumber\\
&&-\,\,4\,U_{\ell'\,1}\,U_{\ell\,1}\,U_{\ell'\,2}\,
U_{\ell\,2}\,\sin^2\left(\varphi^0_{2\,1}\right){\Big]}
\quad.\label{eq:modd}
\end{eqnarray}
The first term on the rhs of the above equation
is the contribution
from the non--relativistic mass eigenstate
 to the 
$P_{\ell\ell^\prime}\left(E,L,\{\xi_k\}\right)$ at the $\nu_{\ell^\prime}$
event
at time $t_{II}$ while the second term is the
contribution  from the relativistic mass eigenstates
 to the  
$\nu_{\ell^\prime}$
event at the earlier time $t_I$.
Contained in the above expression is the 
the fundamental ``collapse of the wave packet'' postulate of the 
orthodox interpretation of the quantum mechanics.
That is, given a single $\nu_\ell$ emitted at the source, 
if the event occurs at $t_I$ no event occurs at $t_{II}$, and vice versa.

There are several observations that one may make about the result
(\ref{eq:modd}). These observations follow.

{\em First:}

\begin{quote}
The neutrino oscillation probability now  contains only one length scale,
$\xi_1=\Delta m^2_{2 1}$, 
 $\xi_2=\theta$, $\xi_3=\beta$, 
and $\xi_4=\psi$.
However, this loss of length scale is related to a manifestly
different expression for neutrino oscillation probabilities.
\end{quote}

{\em Second:}

\begin{quote}
If $m_3$ is in the range 
of a fraction of an MeV to a few tens of MeV, one cannot base the analysis
of existing neutrino oscillation data in terms of Eq. (\ref{eq:d}), or
Eq. (\ref{eq:modd}), alone. 
For the zenith--angle dependence of the  atmospheric neutrino anomaly data
\cite{Kamioka}
neutrino energies in the GeV range
are involved. This meets the requirements under which
Eq. (\ref{eq:d}) is operative.
On the other hand, the physical situation for part of the
LSND events (energy 
range between 20 MeV and Michel spectrum cutoff of 52.8 Mev \cite{LSND}),
and the reactor experiments (average $\overline{\nu}_e$ energy about 5 MeV 
\cite{R1,R2,R3,R4}) (\ref{eq:modd}), the physical conditions for
Eq. (\ref{eq:modd}) may be satisfied. 

\end{quote}

{\em Third:}

\begin{quote}
For the solar--neutrino deficit one may speculate that $m_3$ may be such that
an energy--dependent transition takes place from  the non-relativistic regime
to relativistic regime, thus accounting for the apparent energy dependence
of the solar--neutrino deficit. So, consider a situation where the length
scales are such that all $\sin^2\left(\varphi^0_{\jmath\imath}\right)$
average to $1/2$. Then
\[
P_{ee}\left[\mbox{Eq.}\,\,(\ref{eq:d})\right]
\,=\,1\,-\,2\,U^2_{e1}\,U^2_{e2}
\,-\,2\,U^2_{e1}\,U^2_{e3}
\,-\,2\,U^2_{e2}\,U^2_{e3}\quad,
\]
\[
P_{ee}\left[\mbox{Eq.}\,\,(\ref{eq:modd})\right]
=U^4_{e1} \,+\,U^4_{e2} \,+\,U^4_{e3}\quad.
\]
Exploiting unitarity of the mixing matrix I immediately see that
above speculation, within the defined context, has no consequence because
\[
\mbox{Unitarity}\,\,\Rightarrow\,\,
P_{ee}\left[\mbox{Eq.}\,\,(\ref{eq:d})\right] \,=\,
P_{ee}\left[\mbox{Eq.}\,\,(\ref{eq:modd})\right]\quad.
\]

\end{quote}

{\em Fourth:}

\begin{quote}
In astrophysical environments, such as for neutrinos observed in 
the supernova 1987a \cite{1987a,MRoos}, 
it may happen that $m_3$ is such that the evolution
of $\vert \nu_3\rangle$
envelope does not escape the astrophysical environment.
That is, $p/m$ associated with the envelope of
$\vert \nu_3\rangle$ is less than the escape velocity
\begin{equation}
\mbox{Non Relativistic:}\quad E \,<\, m\,+\,{{r_g}\over {2\, r}}\quad,
\end{equation}
where $r_g\equiv 2\, G M$ is the gravitational radius 
of the astrophysical object.
Under these
circumstances Eq. (\ref{eq:modd}), for a detector at Earth, reduces to:
\begin{eqnarray}
&&P_{\ell\ell^\prime}\left(E,L,\{\xi_k\}\right)
= 
\left(U_{\ell^\prime 3}\,U_{\ell 3}\right)^2 
\,\Theta\left(E \,-\, m\,+\,{{r_g}\over {2\, r}}\right)\nonumber\\
&&\,+\,
\left(
U_{\ell^\prime 1}\,U_{\ell 1}
\,+\,
U_{\ell^\prime 2}\,U_{\ell 2}\right)^2 
-\,\,4\,U_{\ell'\,1}\,U_{\ell\,1}\,U_{\ell'\,2}\,
U_{\ell\,2}\,\sin^2\left(\varphi^0_{2\,1}\right)
\,.\label{eq:moddd}
\end{eqnarray}
In the above expression
$\Theta(\cdots)$ is the usual step function, vanishing for its argument
less than zero and equal to unity for its argument greater or equal
to unity.

\end{quote}

Now I rewrite Eq. (\ref{eq:modd}) in a form that makes the deviations
of result (\ref{eq:modd}) explicit from the corresponding two--flavor scenario
with two relativistic mass eigenstates.  This new form of
Eq. (\ref{eq:modd}) reads:
\begin{equation}
P_{\ell\ell^\prime}\left(E,L,\{\xi_k\}\right)
= 
\Delta_{\ell\ell^\prime}
\,-\,A_{\ell\ell^\prime}\,\sin^2\left(\varphi^0_{2\,1}\right)
\quad,\label{eq:nf}
\end{equation}
with
$ \Delta_{\ell\ell^\prime}\,=\,
\left(U_{\ell^\prime 3}\,U_{\ell 3}\right)^2 \,+\,
\left(
U_{\ell^\prime 1}\,U_{\ell 1}
\,+\,
U_{\ell^\prime 2}\,U_{\ell 2}\right)^2 $, or more explicitly
\begin{equation}
\Delta=
\delta+
\left(
\begin{array}{ccccc}
\mbox{s}^4_\beta+\mbox{c}^4_\beta -1 &{\,\,}&
2\mbox{c}^2_\beta \mbox{s}^2_\beta \mbox{s}^2_\psi &{\,\,}&
2\mbox{c}^2_\beta \mbox{s}^2_\beta \mbox{c}^2_\psi \\
2\mbox{c}^2_\beta \mbox{s}^2_\beta \mbox{s}^2_\psi &{\,\,}&
2 \mbox{c}^2_\beta \mbox{s}^2_\psi(\mbox{c}^2_\beta \mbox{s}^2_\psi-1) &{\,\,}&
2\mbox{c}^4_\beta \mbox{c}^2_\psi\mbox{s}^2_\psi \\
2\mbox{c}^2_\beta \mbox{s}^2_\beta \mbox{c}^2_\psi &{\,\,}&
2\mbox{c}^4_\beta \mbox{c}^2_\psi\mbox{s}^2_\psi&{\,\,}&
2 \mbox{c}^2_\beta \mbox{c}^2_\psi(\mbox{c}^2_\beta \mbox{c}^2_\psi-1)
\end{array}
\right)\,,\label{eq:delta}
\end{equation} 
and
\begin{equation}
A_{\ell\ell^\prime}\equiv 4\,U_{\ell'\,1}\,U_{\ell\,1}\,U_{\ell'\,2}\,
U_{\ell\,2}
\quad.
\end{equation} 
In Eq. (\ref{eq:delta}) $\delta$ is a $3\times 3$ identity matrix.

Only when both $\beta$ and $\psi$ vanish does Eq. (\ref{eq:nf})
reduce to the expression for
the two flavor scenario
with two relativistic mass eigenstates --- for then 
$U(\theta,\,\beta,\,\psi)$ becomes block diagonal with no mixing with 
$\vert \nu_3\rangle$, as
\begin{equation}
\beta=\psi=0\,:\quad
\Delta =
\left(
\begin{array}{ccc}
1 & 0 & 0\\
0 & 1 & 0\\
0 & 0 & 1
\end{array}
\right)\,,\quad
A =
\left(
\begin{array}{ccc}
\mbox{s}^2_{2\theta} & -\,\mbox{s}^2_{2\theta} & 0 \\
-\,\mbox{s}^2_{2\theta} & \mbox{s}^2_{2\theta} & 0\\
0 & 0 & 0
\end{array}
\right)\quad.
\end{equation}

It is, therefore, concluded that there is enough richness in the 
three flavor neutrino oscillation phenomenology that unless an argument can
be made that this structure is incapable of accommodating the existing 
neutrino oscillation data
there is no necessity to put forward the sterile neutrino hypothesis.
I hasten to add that the purpose of this paper was not to show that the
uncovered structure and length scales can necessarily accommodate the existing and
the forthcoming data, but only to reveal the often ignored structure and the
hidden complexity of the 
three--flavor neutrino oscillation phenomenology.

\section{Conclusions}

In these notes I  presented  a 
 critique of the standard three--flavor neutrino oscillation 
framwork.
I have argued that the standard three--flavor neutrino oscillation
framework has rich structure that must not be ignored in making
various approximations and designing  experiments. Specifically,
the design proposal of the MINOS at Fermilab based
on a two--mass eigenstate framework may require  serious 
revision if there is strong mixing between all three flavors of neutrinos.
The general lesson to learn here is that:
\begin{quote}
Even if one is searching
for neutrino oscillations between two flavors it may not be advisable
to work in terms of the well known formula (cf, \cite[p. 17]{MatF}):
\[
P(\nu_a\rightarrow\nu_b) = \sin^2(2\theta)\,\sin^2\left(1.27{ {\Delta m^2 L}
\over {E} }\right)\quad,
\]
where $\Delta m^2$ is measured in $\mbox{eV}^2$, $L$ in meters (kilometers),
and $E$ in $\mbox{MeV}\,\, (\mbox{GeV})$.
\end{quote}
Detailed analytical and graphical analysis of LSND NOE revealed
the interesting roles that various relative phases play in neutrino
oscillation probabilities. For example, in the context of LSND NOE and
KARMEN,
the amplitude ${\cal{A}}_{3(2,1)}$,
defined in Eq. (\ref{eq:zzz}), vanishes for certain
values of mixing angles as a result of opposite signs of two equal and opposite
contributions. \footnote{Such cancellations are a generic feature of 
three, or more,  state systems  and range from
neutrino oscillation phenomenology to laser 
oscillation without  population inversion \cite{GGP}. }
I further argued, for instance, why the canonical wisdom
of ``one mass scale dominance'' may fail in certain 
unexpected sectors of the neutrino--oscillation parameter space.
By comparing the solar neutrino deficit and the results on
$\overline{\nu}_e$ disappearance in the reactor experiments I
reached the conclusion that the standard three--flavor neutrino oscillation
framework requires one of the mass--squared differences to be
$\ll$ $10^3\,\,\mbox{ev}^2$. This, when
coupled with the very recent astronomical observations,
left open the possibility that
one of the mass eigenstates may be non--relativistic and hence introduce
a fundamental modification to the standard 
neutrino oscillation phenomenology
without invoking a sterile neutrino. 
If the transition from the 
non-relativistic to the relativistic regime happens for energies relevant
to the Reactor and LSND NOE then
one must consider an {\em ab intio} analysis of the existing data.

{\bf Acknowledgements}

I extend my  thanks to Drs. Christoph Burgrad,
Terry Goldman, Peter Herczeg, Mikkel Johnson,
Bill Louis, John McClelland, Ion Stancu,  Hywel White, and Nu Xu
for our continuing conversations  on neutrino oscillations and 
other matters of physics. 

{\em This work was done, in part, under the auspices of the 
U.S. Department of Energy.}

\newpage

\begin{appendix}

\section{Comments and Observations on Eqs. (1) and (2) and
A Brief Outline of the Standard
Neutrino--Oscillation 
Phenomenology}

Equations (\ref{eq:dnew}) and (\ref{eq:d})
 require several comments and observations.
These remarks follow.

While analyzing experimental situations, the integral appearing in
Eq. (\ref{eq:d}) is evaluated by summing over  
approximately--constant bins of $f_\ell(E)$.
In this context I note that
the integral
\begin{equation}
I_{\jmath\imath}=
\int_{E_{min}}^{E_{max}} dE \,
\sin^2\left(2\,\pi \,{L\over\lambda^{\rm osc}_{\jmath\imath}}\right)
=
\int_{E_{min}}^{E_{max}} dE \,
\sin^2\left( { {C_{\jmath\imath}} \over E}\right)\,,
\end{equation}
with $C_{\jmath\imath}=\alpha L \Delta m^2_{\jmath\imath}\,$ may
be  evaluated by introducing $y\equiv 1/E$, and using the result
\begin{eqnarray}
\int_{E_{min}}^{E_{max}} dE \,
&&\sin^2\left( { {C_{\jmath\imath}} \over E}\right) =
\int_{y_1}^{y_2}\left( -\, { {dy}\over{y^2} }\right)
 \sin^2\left(y\,C_{\jmath\imath}\right)
\nonumber\\
&&
= -\,{ {\cos(2\,y_2\,C_{\jmath\imath})}  \over {2\,y_2} }
\,- \, 
{\mbox{Si}}(2\,y_2\,C_{\jmath\imath})\,C_{\jmath\imath} \,+\,{1\over {2\,y_2}}
\nonumber\\
&&\quad+\,{{ {\cos(2\,y_1\,C_{\jmath\imath})} } \over {2\,y_1} }
\,+ \, 
{\mbox{Si}}
(2\,y_1\,C_{\jmath\imath})\,C_{\jmath\imath} \,-\,{1\over{2\,y_1}}\quad.
\end{eqnarray}

Referring to Eq. (\ref{eq:d}),
it is important to note  how it arises.
Apart from the already indicated assumptions,
the fundamental assumption of the neutrino--oscillation 
phenomenology, contained in Eq. (\ref{eq:d}),
is that weak flavor eigenstates for
$\nu_e,\nu_\mu,\nu_\tau$, and  $\overline\nu_e,\overline\nu_\mu,
\overline\nu_\tau$, are not mass eigenstates. Why this is so, no 
one knows at the present time and the question remains a deep 
mystery. To appreciate the nature of this mystery, note that
from
the 1939 classic paper of Wigner \cite{EPW1939} we have learned that
every physical state
is described by two Casimir invariants associated
with the Poincar\'e group of space--time symmetries. These Casimir invariants
are directly connected to {\em mass} and {\em spin}. Why, therefore, do
weak interactions  and space--time symmetries intermingle in such a manner as
to prefer linear {\em superpositions} of mass eigenstates? 
I shall return to this
subject elsewhere building on  some recent
work on the subject \cite{DVAnp,VVD,PCP,PCP1}.
 
Here is a brief outline of the standard
neutrino--oscillation 
phenomenology. 
Assume that in the ``creation region,''
${\cal R}_c$, located at ${\vec r}_c$,
 a weak eigenstate with energy $E$
(with appropriate uncertainty dictated not only by
the Heisenberg Principle but also determined by the masses associated with mass
eigenstates) denoted by
$\vert\nu_\ell,\,{\cal R}_c\rangle$, is produced at ${\vec r}_c$,
with the clock set to $t=0$. Each of the three neutrino mass eigenstates
shall be represented by $\vert \nu_\imath \rangle$; $\imath=1,2,3$. 
So I have the linear superposition:
\begin{equation}
\vert \nu_\ell, \,{\cal R}_c\rangle =
\sum_{\imath=1,2,3} U_{\ell\imath}\,\vert\nu_\imath\rangle\quad,
\label{nu0}
\end{equation} 
where $\ell=e,\mu,\tau$ represent the  weak flavor 
 eigenstates (corresponding
to electron, muon, and tau 
neutrinos, respectively).
The $\vert \nu_1\rangle$, $\vert \nu_2\rangle$, and 
$\vert \nu_3\rangle$ correspond to the three mass eigenstates
of masses $m_1,\, m_2,\,\,\mbox{and}\,\, m_3$, respectively. Under the 
already--indicated assumptions the unitary mixing matrix $U_{\ell\imath}$
may be parameterized by three angles
and is given by Eq. (\ref{eq:umix}).

At a time ``$t$''$ =t > 0$,
I wish to study the weak flavor eigenstate in the ``detector region,''
${\cal R}_d$, located at ${\vec r}_d$.
Under the above indicated assumptions the neutrino evolution is given by the
expression 
\begin{equation}
\vert{\cal R}_d\rangle =
\exp\left(-\,{i\over {\hbar }} \int_{{t}_c}^{{t}_d}
H \mbox{dt} \,+\, {i\over \hbar}
\int_{{\vec r}_c}^{{\vec r}_d}
\vec P\cdot d\vec x
\right)
\vert\nu_\ell,\,{\cal R}_c\rangle\quad.
\end{equation}
Here $H$ is the time translation operator, the Hamiltonian,
 associated with  the system; $\vec P$ is the operator
for spatial translations, the momentum operator, and 
$\left[H(t,\vec x),\,{\vec P}(t,\vec x)\right]=0$. It shall be noted
that
$\vert\nu_\ell,\,{\cal R}_c\rangle$ has evolved to  a state
$\vert{\cal R}_d\rangle $ that, in general, is {\em not}
an eigenstate associated with $\nu_e$, $\nu_\mu$, or $\nu_\tau$. Instead,
it is a linear superposition of states associated with 
$\nu_e$, $\nu_\mu$, or $\nu_\tau$.
The ``neutrino oscillation probability'' from a state
$\vert\nu_\ell,\,{\cal R}_c\rangle$ 
to another state $\vert \nu_{\ell^\prime}, \,{\cal R}_d\rangle$
is now obtained by calculating the projection
$\langle\nu_{\ell'},\,{\cal R}_d\vert{\cal R}_d\rangle $,
i.e., the amplitude for 
$\vert \nu_\ell,\,{\cal R}_c\rangle \rightarrow
\,\vert\nu_{\ell'},\,{\cal R}_d\rangle$, and then multiplying it by its
complex conjugate. An 
algebraic exercise that exploits (i)
the unitarity of the neutrino mixing matrix
$U(\theta,\,\beta,\,\psi)$, (ii) orthonormality 
of the mass eigenstates, (iii) certain trigonometric identities, 
makes (iv)
 the standard observation
(for neutrino oscillations) that $(E_\jmath-E_\imath) 
=\Delta m_{\jmath\imath}^2 c^4/ (2 E)$, where $\Delta m_{\jmath\imath}^2
\equiv m_\jmath^2-m_\imath^2$ and $E_\jmath\simeq E_\imath \equiv E$,
and (v) sets $\vert {\vec r}_d - {\vec r}_c\vert = L$,
yields Eq.  (\ref{eq:d}) for
$P_{\ell\ell^\prime}\left(E,L,\{\eta_k\}\right)$.

\section{A Simple Theorem  on the  Inverted Mass Hierarchy}

\begin{quote}
{ Under}

\begin{equation}
\Delta m^2_{2 1} \longleftrightarrow \Delta m^2_{3 2}\quad,\label{eq:i1}
\end{equation}

{ and simultaneous change in }
$U(\theta,\beta,\psi)$ { of the form}

\begin{equation}
U_{\ell'\,1}\,U_{\ell\,1} \longleftrightarrow 
\pm\,
U_{\ell'\,3}\,U_{\ell\,3}\quad,\label{eq:i2}
\end{equation}

\begin{equation}
U_{\ell'\,2}\,U_{\ell\,2} \longleftrightarrow 
\pm\, 
U_{\ell'\,2}\,U_{\ell\,2}\quad,\label{eq:i3}
\end{equation}
{
the ${\cal P}(\nu_\ell \rightarrow\nu_{\ell'})$,
given by Eq. (\ref{eq:dnew}),
remains unchanged.}
\footnote
{Either the upper $+$ sign, or the lower $-$ sign, in {\em both}
of the above equations
is meant to be taken in any given transformation.
}
\end{quote}

The invariance of various neutrino oscillation probabilities
${\cal P}(\nu_\ell \rightarrow\nu_{\ell'})$, under the above defined
change in neutrino--oscillation parameters,
does not imply that other physical phenomena too remain unaffected
under the change in the neutrino--oscillation parameters defined above.

For considerations on the evolution of the universe, and formation
of the large--scale structures in the universe, it may be noted
that given a mass eigenstate configuration $m_3 \gg m_2\ge m_1$
(by definition, the ``standard mass hierarchy''), the corresponding
 ``inverted mass hierarchy''
(by definition,  $m_3 \ge m_2\gg m_1$) defined by the above symmetry,
{\em necessarily\/} has a larger $\sum_{\imath} m_\imath$ than
the $\sum_{\imath} m_\imath$ associated with the (corresponding) 
standard mass hierarchy.\footnote{
Similar comments apply in the context of
gravitational effects on  physical phenomenon involving neutrinos
in astrophysical environments.
An exception is the  
gravitationally induced modification to the  neutrino--oscillation 
probabilities, where a similar symmetry as mentioned above exists \cite{AB}.}

Within the above framework (i.e., three flavor neutrino oscillation
phenomenology with Dirac neutrinos without CP violation), 
the  neutrino--oscillation experiments cannot distinguish
between the standard mass hierarchy and the inverted mass hierarchy.
Additional physical phenomenon must be considered to distinguish
the standard and inverted mass hierarchies.

\section{Energy Conservation and Neutrino Oscillations}

This section is included here purely for 
pedagogic reasons, and may be skipped by the ``experts.''

Let us begin with posing a pedagogic question. To keep the 
algebraic details simple, I will 
confine to a situation where the mixing angles
$\beta$ and $\psi$ vanish. Then, consider an electron neutrino in the creation 
region:
\begin{equation}
\vert \nu_e,\, {{\cal R}_c}\rangle = c_\theta \vert \nu_1\rangle
+s_\theta \vert \nu_2\rangle \quad, \label{z1}
\end{equation}
and, again for the sake of argument, 
I will confine to a situation where the neutrino mass eigenstates
have same momenta, and hence different energies. The expectation of value
of energy
in the state given by (\ref{z1}) is
\begin{equation}
\langle E_{\nu_e} \rangle = 
c_\theta^2\,E_1 + s_\theta^2\,E_2\quad.\label{z2}
\end{equation}
The state $\vert \nu_e,\, {{\cal R}_c}\rangle$ evolves with time, and at any
given time it has a definite probability of being found (on measurement)
as an electron neutrino, and one minus that probability of being found
as a muon neutrino. The energy expectation value of the energy 
remains constant in time. But now suppose I make a measurement and detect
a muon neutrino. The linear superposition of the electron and muon neutrino
state now collapses to muon neutrino
\begin{equation}
\vert \nu_\mu,\, {{\cal R}_d}\rangle = - s_\theta \vert \nu_1\rangle
+c_\theta \vert \nu_2\rangle \quad. \label{z3}
\end{equation}
The expectation value of the energy
is now
\begin{equation}
\langle E_{\nu_\mu} \rangle = 
s_\theta^2\,E_1 + c_\theta^2\,E_2\quad.\label{z4}
\end{equation}
The posed question is: Where did the excess energy,
\begin{equation}
\Delta E_{excess} \equiv \langle E_{\nu_\mu} \rangle
-\langle E_{e} \rangle = c_{2\theta} \left(E_2-E_1\right)\quad, \label{z5}
\end{equation}
come from?

To answer this question it should be realized that in order
for  the detector to detect a  muon neutrino it must be 
able to make a measurement over a period \footnote{In this section I shall
not set $\hbar$ and $c$ equal to unity.}
\begin{equation}
\Delta t \approx {c\over{\lambda^{\rm osc}/4}} = {{\hbar E} \over 
{{\Delta m^2_{2\,1}}} c^4}, \qquad E=c_\theta^2 E_1 + s_\theta^2 E_2\quad.
\end{equation}
As such, the act of measurement must impart a minimum energy (via
the scattering process in the detector) $\Delta E_{Heisenberg}= 
\approx\hbar/(2 \Delta t)$ to the detected 
muon neutrino. It is readily found that
\begin{equation}
\Delta E_{Heisenberg} = {{\Delta m^2_{21} c^4}\over {2\,E}}\quad.
\end{equation}
Referring to Eq. (\ref{z5}), now note that 
$E_2-E_1\approx \Delta m^2_{2\,1} c^4/ (2\,E)$. 
Thus
\begin{equation}
\Delta E_{excess}  = c_{2 \theta}\, \Delta E_{Heisenberg}\quad.
\end{equation}
So what at first might have appeared as a violation of energy conservation
is precisely the energy, or more accurately the
$c_{2\theta}$ fraction of $\Delta E_{Heisenberg} $,
 that the act of measurement imparts to the
detected muon neutrino. 

For a more formal discussion of this point see Ref. \cite{TG}.

\end{appendix}


\newpage

\begin{thebibliography}{999} 


\bibitem{DVAnp}
D. V. Ahluwalia,  {\em Int. J. Mod. Phys. A\/}, {\bf 11}, 1855 (1996).



\bibitem{DR1}
S. P. Mikheyev and A. Yu. Smirnov, {\em Sov. J. Nucl. Phys.}
{\bf 42}, 913 (1985);\\
 L. Wolfenstein, {\em Phys. Rev. D\/} {\bf 17}, 2369 (1978).

\bibitem{AB}
D. V. Ahluwalia and C. Burgard, {\em Gen. Rel. and Grav.\/} 
{\bf 28}, 1161 (1996).
A continuing debate exists on the subject of
``gravitationally induced neutrino oscillation phases.''
In our opinion, our reply and the preprint of Y. Grossman and H. Lipkin
is sufficient to resolve the confusion. The relevant references are:
T. Bhattacharya, S. Habib and E. Mottola, gr-qc/9605074;
D. V. Ahluwalia and C. Burgard, gr-qc/9606031;
Y. Grossman and H. Lipkin, gr-qc/9607201;
C. Y. Cardall and G. M. Fuller, hep-ph/9610494;
N. Fornengo, C. Giunti, C. W. Kim, and J.Song, hep-ph/9611231.
The basic result of our work can be summarized as follows:
The phenomenon of neutrino oscillations provides a ``flavor oscillation clock,''
and this clock red shifts as required by Einstein's general relativity.




\bibitem{PRW}
D. P\'{\i}riz, M. Roy, and J. Wudka, {\em Phys. Rev. D\/} {\bf 54}, 2761 (1996).




\bibitem{KPbook}
C. M. Kim and A. Pevsner, {\em Neutrinos in Physics and Astrophysics\/}
(Harwood Academic Publishers, 1993).

\bibitem{BK} B. Kayser (with F. Gibrat--Debu and F. Perrier),
{\em The Physics of Massive Neutrinos\/}, (World Scientific, Singapore, 1989).

\bibitem{MPbook} R. N. Mohapatra and P. B. Pal,
{\em Massive Neutrinos in Physics and Astrophysics\/}
(World Scientific, Singapore, 1991).


\bibitem{HL}
H. J. Lipkin, {\em Phys. Lett. B\/} {\bf 348}, 604 (1995).

\bibitem{TG}
T. Goldman, {\em 
 hep-ph/9604357}.

\bibitem{ABIV}
E. Alfinito, M. Blasone, A. Iorio, and G. Vitiello, {\em Phys. Lett. B\/}
{\bf 362}, 91 (1995); Also see, LANL preprint archive: hep-ph/9601354.


\bibitem{Kamioka} 
Y. Fukuda et al., {\em Phys. Lett. B\/}  {\bf 335},237 (1994);
\\ For a comment on the statistical significance of the
quoted zenith-angle dependence of the atmospheric neutrino
anomaly by Fukuda et al., 
see D. Saltzberg, {\em Phys. Lett. B\/} {\bf 355}, 499 (1995);
\\
B. C. Barish, {\em Nucl. Phys. B (Proc. Suppl.)\/} {\bf 38}, 343 (1995).


\bibitem{tw}
J. G. Learned, S. Pakvasa, and T. J. Weiler, {\em Phys. Lett. B\/}
{\bf 207}, 79 (1988).



\bibitem{DVA21}
D. V. Ahluwalia and T. Goldman, forthcoming.



\bibitem{MR} 
L. Maiani, {\em Phys. Lett. B\/} {\bf 62}, 183 (1976). \\
However, the manner
in which rows and columns are labeled and notation for angles is
essentially that
of Ref. \cite[Eq. 6.21]{KPbook}.



\bibitem{LSND}
C. Athanassopoulos et al., {\em Phys. Rev. Lett.\/}
{\bf 75}, 2650 (1995);\\
C. Athanassopoulos {et al.}, {\em Phys. Rev. C} (1996, submitted);
LANL archive preprint: nucl-ex/9605001.




\bibitem{KARMEN}
B.\ Bodmann {\it et\ al.\/}, {\em Phys. Lett. B\/} {\bf 267}, 321 (1991);
B.\ Bodmann {\it et\ al.\/}, {\em Phys. Lett. B\/} {\bf 280}, 198 (1992);
B.\ Zeitnitz {\it et\ al.\/}, {\em Prog. Part. Nucl. Phys.\/}
 {\bf 32}, 351 (1994).


\bibitem{BL} 
W. C. Louis,  (private communication, April 1996).

\bibitem{Fogli} 
G. L. Fogli, E. Lisi, and G. Scioscia, 
{\em Phys. Rev. D\/} {\bf 52}, 5334 (1995); \\
 G. L. Fogli and E. Lisi, {\em Phys. Rev. D\/} {\bf 52}, 2775 (1995).

\bibitem{R1}
G. Zacek {et al.,\/} {\em Phys. Rev. D\/} {\bf 34}, 2621 (1986).

\bibitem{R2}
G. S. Vidyakin {et al.,\/} {\em Pis'ma Zh. Eksp. Teor.  Fiz.\/}
{\bf 59}, 364 (1994) 
 [English translation: {\em JETP Lett.\/} {\bf 59}, 390 (1994)].

\bibitem{R3}
B. Achkar {et al.,\/} {\em Nucl. Phys. B\/} {\bf 434}, 503 (1995).

\bibitem{R4} Z. D. Greenwood {et al.,\/} {\em Phys. Rev. D\/} {\bf 53},
6054 (1996).



\bibitem{Solar} 
Results from SAGE: J. N. Abdurashitov et al.,
{\em Phys. Lett. B\/} {\bf 328}, 234 (1994);\\
Results from GALLEX: P. Anselmann, {\em Phys. Lett. B\/ } {\bf 327}, 377 (1994);\\ 
Results on $^8\mbox{B}$ Solar Neutrinos from Kamiokande II:
K. S. Hirata {et al.}, {\em Phys. Rev. D\/} {\bf 44}, 2241 (1991);\\
Review of the Homestake  Solar Neutrino Experiment
($^{37}\mbox{Cl}$ Experiment): R. Davis, {\em Prog. Part. Nucl. Phys.\/ }
{\bf 32}, 13 (1994);\\
For the Standard Solar Model, and various views
on the status of the
solar neutrino deficit, see, for example:
J. N. Bahcall and M. M. Pinsonneault, {\em Rev. Mod. Phys.\/}
{\bf 64}, 885 (1992);\\
S. Turck--Chie\'ze et al., {\em Phys. Rep.\/} {\bf 230}, 57 (1993);\\
J. N. Bahcall,
{\em Nucl. Phys. B (Proc. Suppl.)\/} {\bf 43}, 41 (1995);\\
S. Turck--Chie\'ze and I. Lopes, {\em Ap. J.\/} {\bf 408}, 347 (1993);\\
T. J. Bowles and V. N. Gavrin, {\em Annu. Rev. Nucl.  Part. Sci.\/} {\bf 43},
117 (1994);\\
S. T. Petcov, {\em Nucl. Phys. B (Proc. Suppl.)\/} {\bf 43}, 12 (1995);\\
K. V. L. Sarma, {\em Int. J. Mod. Phys. A\/} {\bf 10}, 767 (1995);\\
W. Kwong and S. P. Rosen, {\em Mod. Phys. Lett. A\/} {\bf 10}, 1331 (1995); \\
W. C. Haxton, {\em Ann. Rev. Astron. Astrophys.\/} {\bf 33}, 459 (1995);\\
{\it and}
references therein.

\bibitem{GB}
G. B\"orner, {\em The Early Universe: Facts and Fiction}
(Springer--Verlag, Berlin, 1992).



\bibitem{news1}
R. Lieu et al., { Science} {\bf 274}, 1335 (1996);\\
S. Bowyer et al., { Science} {\bf 274}, 1338 (1996).

\bibitem{pc1}
S. Bowyer (private communication, Winter 1996).


\bibitem{news2}
G. Schilling, {Science} {\bf 274}, 1305 (1996); reporting
on ``a pair of papers soon to appear in {\em Astronomy and Astrophysics}''
by H. Arp and H.-D. Radecke.


\bibitem{taumass}
D. Buskulic et al., {Phys. Lett. B} {\bf 349}, 585 (1995).


\bibitem{PDB}
R. M. Barnett et al. (Particle Data Group),
{\em Phys. Rev. D\/} {\bf 54}, 1 (1996).


\bibitem{1987a}
K. Hirata {et al.},  {\em Phys. Rev. Lett.\/}
{\bf 58}, 1490  (1987);\\
R. M. Bionta,
{\em Phys. Rev. Lett.\/}
{\bf 58}, 1494 (1987);\\
M. Aglietta et al.,
in {\em The Standard Model, The Supernova 1987a\/},
J. Tran Thanh Van, ed., Proceedings of the Leptonic Session of the
Twenty-Second Rencontre de Moriond (Editions Fronti\`eres, France,
1987);\\
E. N. Alexeyev  et al.,
in {\em The Standard Model, The Supernova 1987a\/},
J. Tran Thanh Van, ed., Proceedings of the Leptonic Session of the
Twenty-Second Rencontre de Moriond (Editions Fronti\`eres, France, 1987). \\



\bibitem{MRoos} M. Roos,  In {\em The Standard Model, The Supernova 1987a\/},
J. Tran Thanh Van ed., Proceedings of the Leptonic Session of the
Twenty-Second Rencontre de Moriond (Editions Fronti\`eres, France, 1993);\\
J. N. Bahcall,  and S. L. Glashow,  {\em Nature\/} {\bf 326}, 476 (1987). 



\bibitem{MatF} E. Ables et al., The MINOS collaboration,
``P--875: A Long--baseline Neutrino Oscillation Experiment at Fermilab,''
February 1995. 


\bibitem{GGP} 
G. G. Padmabandu {et al.,\/} {\em Phys. Rev. Lett.\/} {\bf 76}, 2053 (1996).



\bibitem{EPW1939}
E. P. Wigner, {\em Ann. of Math.\/}  {\bf 40}, 149 (1939).

\bibitem{VVD}
V. V. Dvoeglazov, {\em Int. J. Theor. Phys.} 34 (1995) 2467.\\
V. V. Dvoeglazov, {\em Nuovo Cim.} 108A (1995) 1467. 

\bibitem{PCP}
D. V. Ahluwalia, {\em P, C, and T Structure of Space--Time} (Kluwer
Academic,   Forthcoming).

\bibitem{PCP1}
D. V. Ahluwalia, M. B. Johnson, and T. Goldman,
{\em Phys. Lett. B\/}  {\bf 316}, 102  (1993).



\end{thebibliography}
\end{document}